\RequirePackage{fix-cm}
\documentclass[smallcondensed]{svjour3} 

 \usepackage{amsmath,amssymb,bm}
\usepackage{floatrow}

 \smartqed 
\usepackage{graphicx}

\begin{document}

\title{Spin Squeezing of Superposition of Biaxial State and two qubit Bell State 
}


\author{Sakineh Ashourisheikhi        
}


\institute{Sakineh Ashourisheikhi        \at
Department of Physics, Yuvaraja's College \\
University of Mysore, Mysore-570005,\,India \\           
\email{samin$ _{-}$ashuri@yahoo .com}       
}


\maketitle
\newpage
\begin{abstract}
 In this study we investigate the spin squeezing in superposition of a Biaxial state [1,2] and Bell state. Numerical and analytical solutions for the length of mean spin, mean
spin direction and spin squeezing are given. It is shown that both the mean spin
direction and spin squeezing parameter are determined by the coefficients of superposition
and the relative phase.      
\end{abstract}

\section{Introduction}
\label{intro}
Spin squeezing is a quantum strategy introduced in 1993 by Kitagawa and Ueda[3] 
which aims to redistribute the fluctuations of two conjugate spin directions among
each other. It has long been known that, spin squeezing [3-5] has attracted considerable attention, both theoretically and experimentally.
The concept of spin squeezing has developed mostly from the study of particle correlations and
entanglement [6-8], as well as the improvement of measurement precision in experiments [4,5,9,10]. Kitagawa and Ueda in Ref. [3] proposed wide studies on squeezing parameters, denoted by $ \xi^{2}_{s} $ which was determined by the well-known photon squeezing. Remarkably, it was realized that spin squeezing is directly connected to a key element in quantum information theory and indicates quantum entanglement[11]. According to definition of Kitagawa and Ueda[3], if a state is spin squeezed there is a relation between spin squeezing parameter $ \xi^{2}_{s} $ and the concurrence [12,13] specifying the entanglement of two spin-half particles.
 Spin squeezing parameters are widely used as measures of many-body correlations because spin squeezing is rather easy to be generated and measured experimentally [14,15]. 

  In this paper, we investigate the mean spin direction, the length of mean spin and spin
squeezing in a superposition of Biaxial state
 \begin{equation}
  \vert\phi \rangle=\frac{\sqrt{2}}{\sqrt{1+cos^{2}\mu}}\left( cos^{2}\frac{\mu}{2}\vert 11  \rangle-sin^{2}\frac{\mu}{2}\vert 1-1  \rangle\right) 
\end{equation} 
[1,2] in the xz-plane and symmetric Bell state. Well known studies to explore spin squeezing of superposition already exist in literature[16,17,18]. In the following, we will derive numerical and analytical solutions for spin squeezing parameters. The spin squeezing parameter is found to be
depend on the superposition coefficients and the relative phases. It is also demonstrated  that spin squeezing exists over a long time with respect to superposition coefficient $ \beta $ for special values of relative
phases. Whereas for other values of relative
phase, existence of spin squeezing is for short duration. Since superposition of quantum states is a fundamental principle of quantum mechanics and also squeezing parameter has close relation to the entanglement of many qubits, this study is essential to a good understanding of quantum mechanics. 

\section{Spin Squeezing of Superposition}
\label{sec:1}
In this section we consider the spin squeezing of a two-qubit state, which is superposed by
a Biaxial state[1,2] and symmetric Bell state $\vert 10  \rangle $ with
 phase  $  \nu$
\begin{equation}
\vert\psi \rangle=\frac{\alpha\sqrt{2}}{\sqrt{1+cos^{2}\mu}}\left( cos^{2}\frac{\mu}{2}\vert 11  \rangle-sin^{2}\frac{\mu}{2}\vert 1-1  \rangle\right)+\beta\,e^{i\nu} \vert 10  \rangle
\end{equation}
where, without loss of generality, the coefficients $  \alpha$ and $  \beta$ are taken to be real  $\alpha^{2}+\beta^{2} =1 $. In order to study
spin squeezing we first consider the mean spin direction and the length of mean spin in the
following section.
\section{Mean Spin Direction and the Length of Mean Spin}
\label{sec:2}
According to the definitions of spin squeezing, we first need to know the mean spin direction
and the length of mean spin which are determined by expectation values $ \langle\hat{J}_{i}\rangle $ with
$  i = \{x, y, z\}$ [19,20]. A straightforward calculation then gives
\begin{equation}
\langle \hat J_{x}\rangle= \frac{2\alpha\beta}{\sqrt{1+cos^{2}\mu}}\,cos\nu \,cos\mu
\end{equation}
\begin{equation}
\langle \hat J_{y}\rangle= \frac{-2\alpha\beta}{\sqrt{1+cos^{2}\mu}}\,sin\nu
\end{equation}
and
\begin{equation}
\langle \hat J_{z}\rangle= \frac{2\alpha^{2}}{1+cos^{2}\mu}\,cos\mu
\end{equation}
From the above, we find that the expectation values $\langle J_{i}\rangle$ depend on the coefficients $ \alpha, \beta $ and
the relative phase $  \nu$. The mean spin direction is in general not along  $\hat{x}, \hat{y}, \hat{z}$ axis. If we denote the mean spin direction by $n_{3}$ and the other two directions perpendicular to
it are denoted by $\hat{n}_{1}  $ and $ \hat{n}_{2} $, respectively, then the directions can be written in spherical coordinates as [19,20,21]
\begin{equation}
\left(\begin{array}{cccc}
\vec{n}_{1} \\

\vec{n}_{2}  \\

 \vec{n}_{3} 
\end{array}\right)\,=\left(\begin{array}{cccc}
-sin\phi & \,\,\,cos\phi & \,\,\,0\\

 -cos\theta\,cos\phi &\,\,\, -cos\theta\,sin\phi &\,\,\, sin\theta \\

 sin\theta\,cos\phi &\,\,\, sin\theta\,sin\phi & \,\,\,cos\theta
\end{array}\right)\,\left(\begin{array}{cccc}
\vec n _{x} \\

\vec{n}_{y}  \\

 \vec{n}_{z} 
\end{array}\right),
\end{equation}
where polar angle $  \theta$
\begin{equation}
\theta=cos^{-1} \left( \frac{\langle J_{z}\rangle}{R}\right)  
\end{equation}

and azimuth angle $  \phi$ is given by

 \begin{equation}
   \phi= 
\begin{cases}
    cos^{-1}(\frac{\langle J_{x}\rangle}{R\, Sin\theta})  ,\,\,& \langle J_{y}\rangle>0,\\
2\pi-cos^{-1}(\frac{\langle J_{x}\rangle}{R\, Sin\theta}),& \langle J_{y}\rangle \leq 0.
\end{cases}
\end{equation}
with
\begin{equation}
R=\sqrt{\langle J_{x}\rangle^{2}+\langle J_{y}\rangle^{2}+\langle J_{z}\rangle^{2}}
\end{equation}
 is the length of mean spin. From the above, we can find that the mean spin direction and
the length of mean spin are determined by the superposition coefficients $\alpha$, $\beta$  and the
phases $\mu$ and $\nu$.\\

  Having studied the mean spin direction, we now consider the length of mean spin. Substituting
$\langle J_{i}\rangle$ in Eq. (9), we have
\begin{equation}
R=\frac{2\alpha}{1+cos^{2}\mu} \sqrt{\beta^{2}cos^{2}\nu cos^{4}\mu+\beta^{2}sin^{2}\nu+cos^{2}\mu}
\end{equation}
which can be calculated numerically. In Fig. 1,
we plot the length of mean spin versus $\beta$ and $\nu$ with different $\mu$.

\begin{figure}
   \begin{minipage}[b]{0.3\textwidth}
      \includegraphics[width=\textwidth,height=3.5cm]{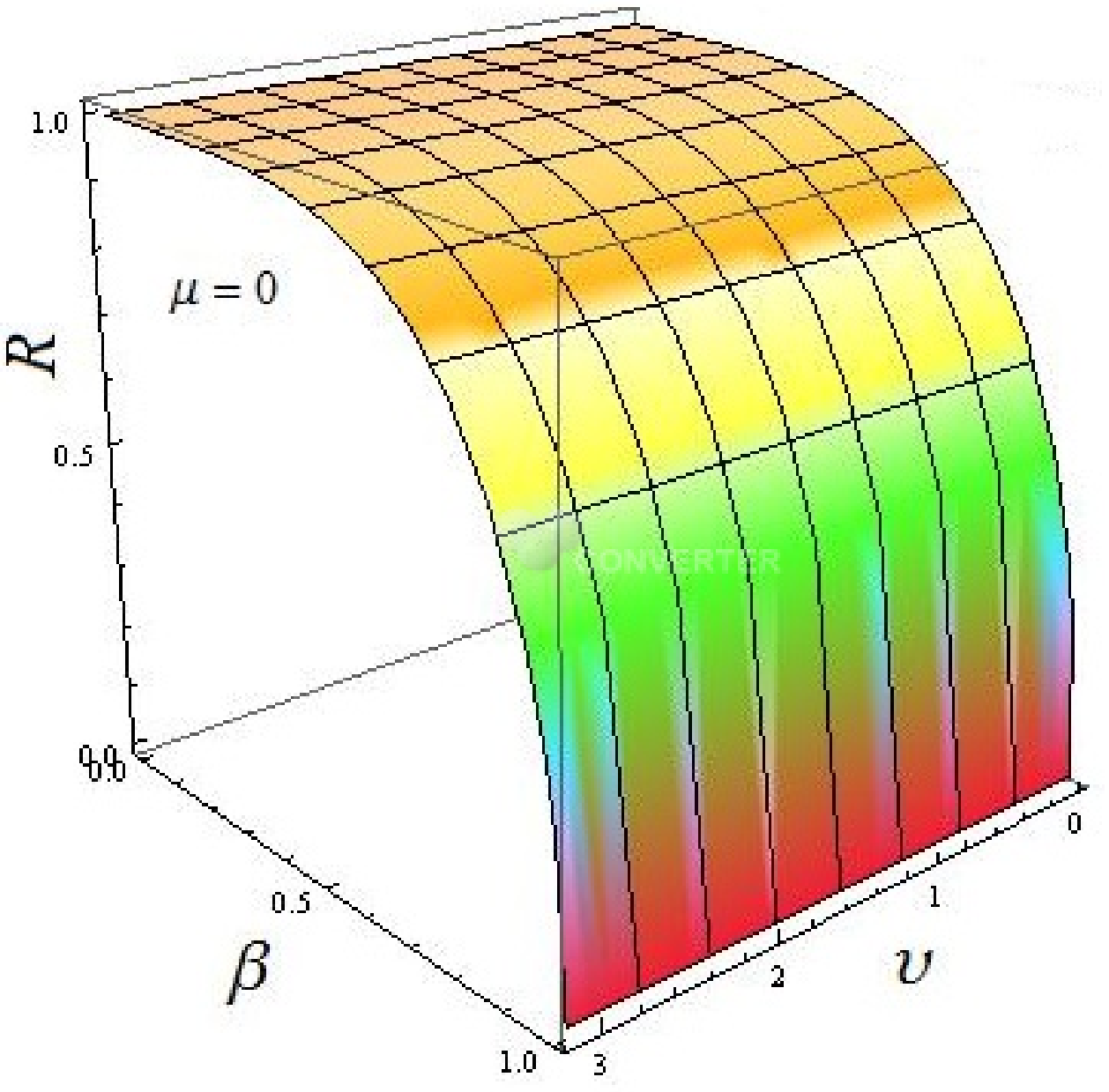}
      
   \end{minipage}
   \hfill
   \begin{minipage}[b]{0.3\textwidth}
      \includegraphics[width=\textwidth,height=3.7cm]{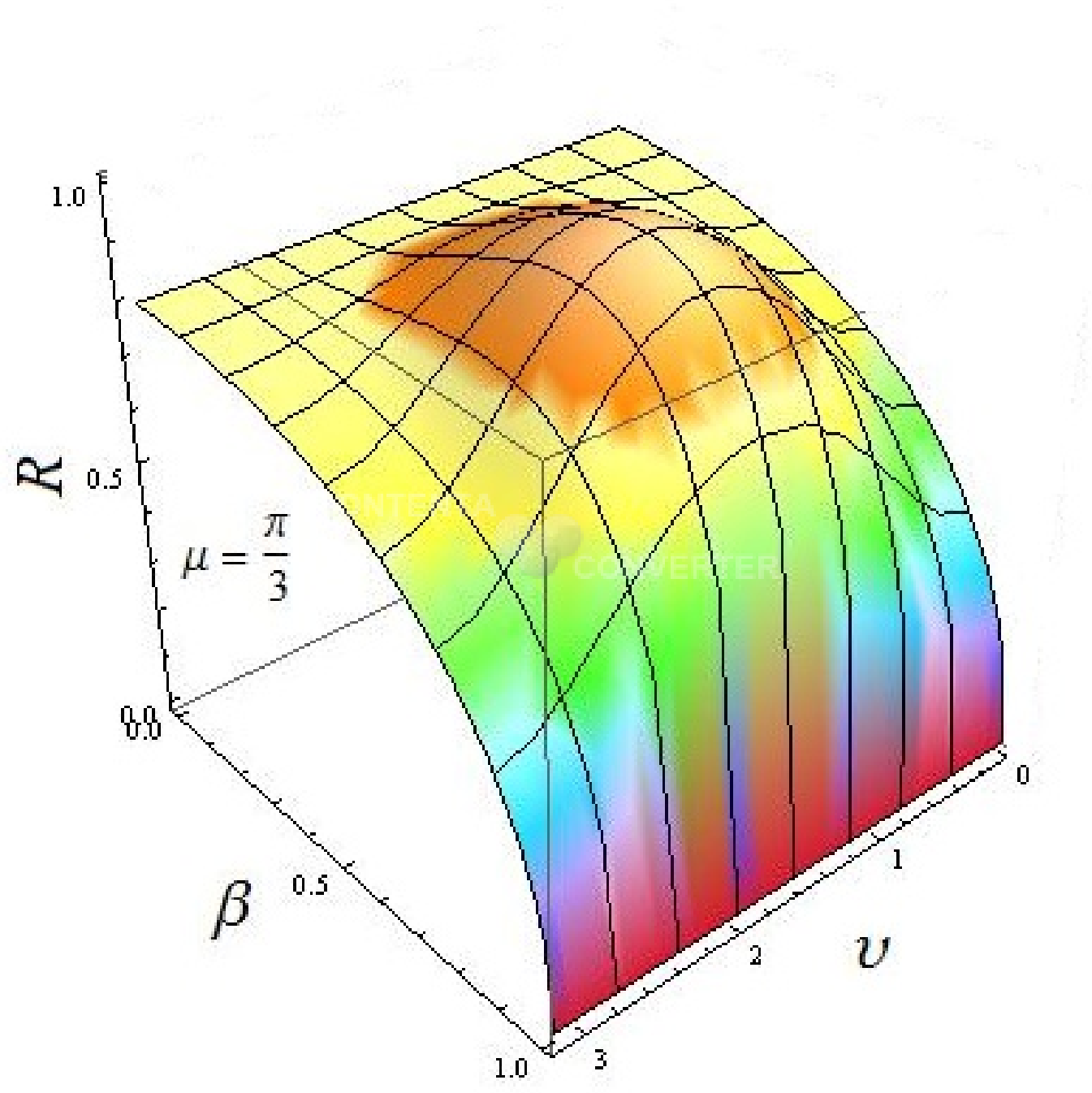}
      
   \end{minipage}
   \hfill
   \begin{minipage}[b]{0.3\textwidth}
      \includegraphics[width=\textwidth,height=3.6cm]{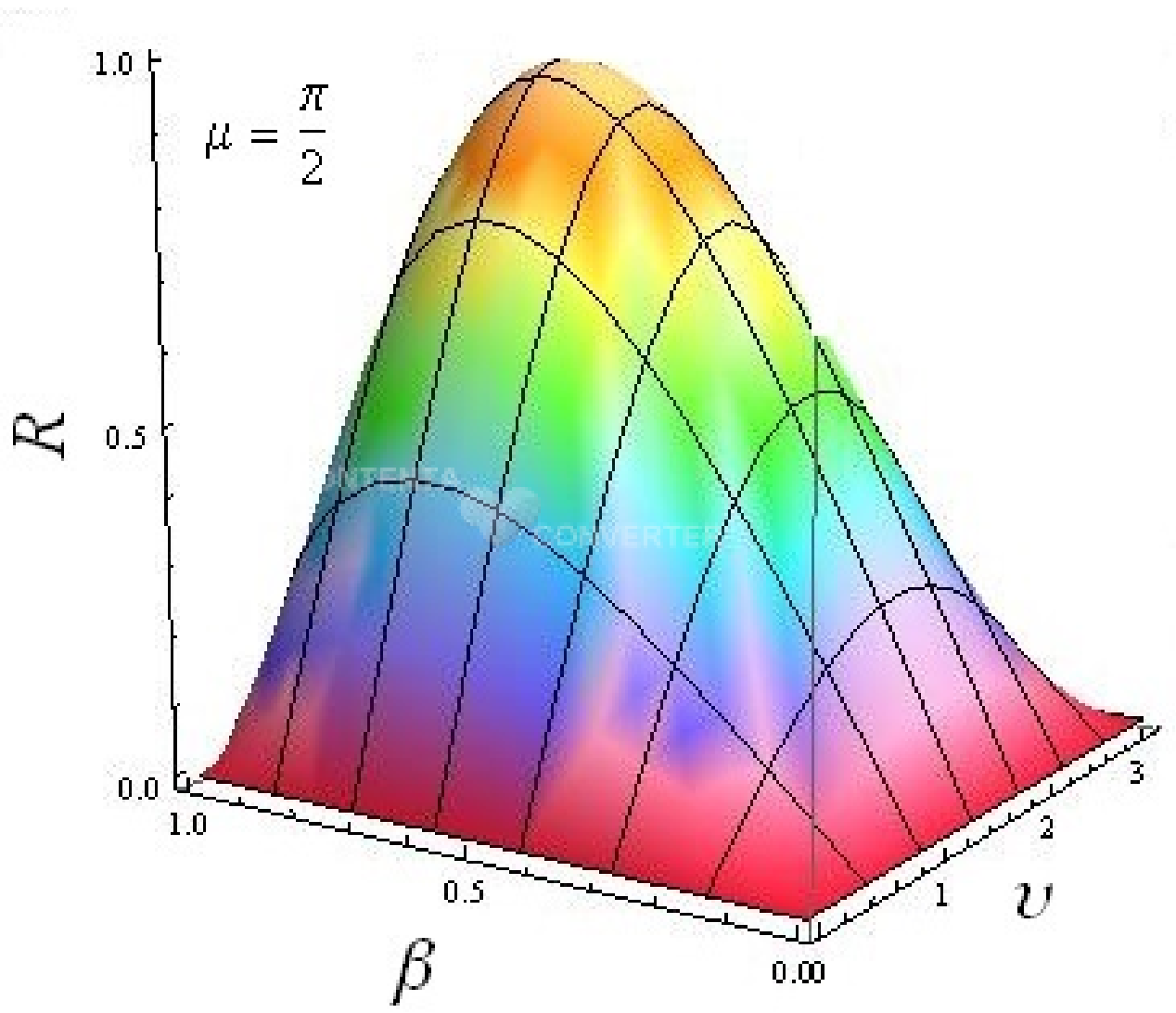}
      
   \end{minipage}
    \caption{The length of mean spin
$  R$ as a function of superposition
coefficient $ \beta $ and phase $ \nu $ for different $ \mu $(Color figure online).
}
\end{figure}

We observe that for $ \mu=0 $, with increasing $ \beta $ the length of mean spin decreases to a minimum. Whereas, for $ \mu=\dfrac{\pi}{3} $ and $ \mu=\frac{\pi}{2} $, with increasing $ \beta $ the length of mean spin first increases to a maximum and then decreases. It is obvious that the relative phases have important effect on the length of the mean spin.
\section{Spin Squeezing}
Knowing the mean spin direction and the length of mean spin we proceed to compute the spin squeezing parameter in order to characterize
the degree of squeezing. Several important
definitions of spin squeezing are proposed by Kitagawa and Ueda [3], Wineland et al. [22]
and Sørensen et al. [23]. In this paper, we employ the criteria of spin squeezing proposed
by Kitagawa and Ueda [3] according to which the degree of spin squeezing is quantified by the following
parameter 
\begin{equation}
\xi^{2}=\frac{4\langle( \Delta{J_{n_{\bot}})^{2}_{min}}\rangle}{N}
\end{equation} 
where $ N=2 $ and
\begin{equation}
( \Delta{J_{n_{\bot}})^{2}_{min}}=\langle J^{2}_{n1}+J^{2}_{n2}\rangle-\sqrt{\langle J^{2}_{n1}-J^{2}_{n2}\rangle^{2}+\langle J_{n1}J_{n2}+J_{n2}J_{n1}\rangle^{2}}
\end{equation}
is the smallest uncertainty of an angular momentum component perpendicular to the mean
angular momentum $ \langle \vec{J}\rangle $ [19,20] with
\begin{equation}
 \langle J_{n1}\rangle=-sin\phi J_{x}+cos\phi J_{y}
\end{equation}
and 
\begin{equation}
 \langle J_{n2}\rangle=-cos\theta cos\phi J_{x}-cos\theta sin\phi J_{y}+ sin\theta J_{z}
\end{equation}
If the inequality $ \xi^{2}<1 $ is satisfied, then the state is said to be a spin squeezed state.
   
   Therefore, we need to calculate the expectation values $ \langle J^{2}_{n1}\rangle $, $ \langle J^{2}_{n2}\rangle $ and $\langle J_{n1}J_{n2}+J_{n2}J_{n1}\rangle$, they which are determined by $ \langle J^{2}_{z}\rangle $, $ \langle J^{2}_{+}\rangle $ and $ \langle J_{+}(2J_{z}+1)\rangle $. A straightforward
calculation then gives
\begin{equation}
\langle J^{2}_{+}\rangle=\frac{-\alpha^{2}\,Sin^{2}\mu}{1+cos^{2}\mu}\,,
\end{equation}
 \begin{equation}
\langle J^{2}_{z}\rangle=\frac{2\alpha^{2}}{1+cos^{2}\mu}(1-\frac{1}{2}Sin^{2}\mu)
\end{equation}
 and
 \begin{equation}
\langle J_{+}(2J_{z}+1)\rangle=\frac{2\,\alpha\,\beta}{\sqrt{1+cos^{2}\mu}}(cos\nu-i\,sin\nu\, cos\mu)
\end{equation}
With the help of the angular momentum relations: $ \langle J^{2}_{x}+J^{2}_{y}\rangle=\frac{N}{2}(\frac{N}{2}+1)- \langle J^{2}_{z}\rangle$, $ \langle J^{2}_{x}-J^{2}_{y}\rangle=Re \langle J^{2}_{+}\rangle $,  $ \langle J_{x}J_{y}+J_{y}J_{x}\rangle=Im \langle J^{2}_{+}\rangle $, $ \langle J_{x}J_{z}+J_{z}J_{x}\rangle=Re \langle J_{+}(2J_{z}+1)\rangle $ and $ \langle J_{y}J_{z}+J_{z}J_{y}\rangle=Im \langle J_{+}(2J_{z}+1)\rangle $, we further obtain the following expectation values:

\begin{align}
\langle J^{2}_{n1}\rangle=\frac{\alpha^{2}}{1+cos^{2}\mu}\left( 1-sin^{2}\mu\,sin^{2}\phi \right)+\beta^{2},
\end{align}

\begin{align}
\langle J^{2}_{n2}\rangle & =\nonumber \frac{-\alpha^{2}\,sin^{2}\mu}{2(1+cos^{2}\mu)}\,cos2\phi\,cos^{2}\theta\\\nonumber
& + \beta^{2}\,cos^{2}\theta+\frac{\alpha^{2}}{1+cos^{2}\mu}(1-\frac{1}{2}sin^{2}\mu)(1+sin^{2}\theta)\\
& + cos\theta\,sin\theta\,cos\phi\,\frac{2\,\alpha\,\beta}{\sqrt{1+cos^{2}\mu}}(sin\nu\,cos\mu-cos\nu),
\end{align}
and  

\begin{align}
\langle J_{n1}J_{n2}+J_{n2}J_{n1}\rangle & \nonumber =-\frac{\alpha^{2}\,sin^{2}\mu}{1+cos^{2}\mu} \\
& - \frac{2\,\alpha\,\beta}{\sqrt{1+cos^{2}\mu}}\,sin\theta\,\left( sin\phi\,cos\nu+ cos\phi\,sin\nu\,cos\mu\right). 
\end{align}

  Substituting these expectation values into Eq. (11) and (12), we can numerically compute spin squeezing parameter. In Fig. 2, we plot spin squeezing parameter $ \xi^{2} $ as function of $ \beta $ for different $ \mu $ and $ \nu $. It shows that in  $ \mu=\frac{\pi}{2} $ and $ \mu=\pi $ for fixed value of $ \nu=\frac{-\pi}{3} $, with increasing $ \beta $, spin squeezing parameter first increases to a maximum and then decreases. whereas, in $\mu=0 $ and $\nu=\frac{-\pi}{3} $, the squeezing parameter decreases to a minimum with increasing of $ \beta $.   
\begin{figure}[h]
\floatbox[{\capbeside\thisfloatsetup{capbesideposition={left,top},capbesidewidth=4cm}}]{figure}[\FBwidth]
{\caption{Spin Squeezing $ \xi^{2} $ as function of $ \beta $ for different $ \mu $ and $ \nu $ (Color figure
online) }\label{fig4.eps}}
{\includegraphics[width=7cm]{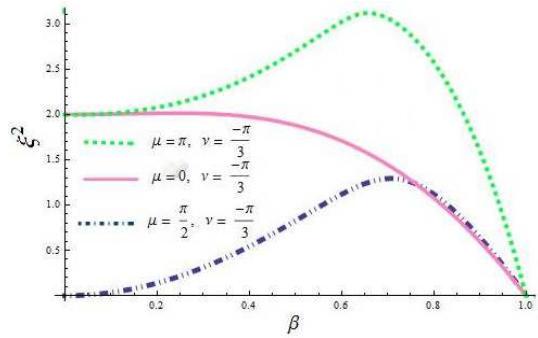}}
\end{figure}

\section{Summary}
We have studied the mean spin direction, lenght of mean spin direction and spin squeezing in a arbitrary superposition of symmetric two qubit Bell state with Biaxial state. Analytical and numerical solutions for the length of mean spin, mean spin direction and spin squeezing are given. The result show that spin squeezing can be generated in superposition of two qubit states. Both the mean spin direction and spin squeezing are sensitive to superposition coefficients and the relative phases. 
The observation shows that in $ \mu=\frac{\pi}{2} $ and $ \nu=\frac{-\pi}{3} $ the spin squeezing exist over a long time with respect to superposition coefficient $ \beta $. Whereas, in $ \mu=0 $ and $ \mu=\pi $ ($ \nu=\frac{-\pi}{3} $) existence of spin squeezing is for short duration.
Since superposition of quantum
states is a fundamental principle of quantum mechanics ,this study enables
a good understanding of quantum mechanics. 




\end{document}